# A displacement-controlled analytical procedure for post-buckling analysis demonstrated by revisiting Euler buckling problem

Xiaguang Zeng[*]

School of Mechatronics Engineering, Foshan University, Guangdong 528000, China

*Corresponding author: xiaguangzeng@lnm.imech.ac.cn; ZengXiaguang@fosu.edu.cn

**Abstract:** In view of the fundamental distinction between the force-controlled model and the displacement-controlled model in buckling problems of structures and the complexity of the asymptotic post-buckling analysis traditionally based on the force-controlled model, alternatively, we provide a straightforward theoretical procedure of the energy method for the buckling and post-buckling analysis completely based on the displacement-controlled model. The Euler buckling behavior is analytically tackled as a static displacement-controlled process as an example of the theoretical procedure, where no force potential energy component but compression and bending strain energy components are considered precisely at the deformed configuration for the total potential energy of the beams. Analytical solutions to the potential energy, structural deformation, internal forces and their critical results are obtained in closed-form for the beams with six typical boundary conditions. We find that these beams have only one unique but universal normalized potential energy surface depending on two dimensionless quantities. The valley bottom pathways on the potential energy surface show that the critical buckling state is not only a bifurcation point but also a valley-ridge inflection point, and the energy increases quadratically before the point and increases linearly with a slope of 2 beyond the point. The axial forces are gradually increasing during post-buckling, but are almost constant, indicating a symmetric neutral bifurcation buckling. Our theoretical expressions generally agree with the counterparts of the force-controlled model but exactly indicate somewhat differences. The present analytical procedure is basically useful to investigate any displacement-controlled buckling problems of beams, plates and shells.

# 1. Introduction

"Everyone loves a buckling problem", as stated by John Hutchinson and Bernie Budiansky in 1979 (Hutchinson and Thompson, 2018), the elastic buckling problems of slender structures as well as thin plates and shells are a classical and charming topic in structure mechanics with a very long history. Focusing on the static Euler buckling of elastic slender structures, it has recently been adopted as an important deformation strategy for stretchable electronics (Su et al., 2012; Xu et al., 2015; Zhao et al., 2021) and many other interesting modern devices, such as the actuators of soft robotics (Aimmanee and Tichakorn, 2018; Chen et al., 2019; Jin et al., 2020), tunable mechanical resonators (Hajjaj et al., 2021; Qiu et al., 2021; Zhang et al., 2017) and sensors (Alcheikh et al., 2021a; El Mansouri et al., 2019), and shock absorbers (Deng et al., 2020; Frenzel et al., 2016; Shan et al., 2015). Such an antiquated mechanics topic, particularly the case under static or quasi-static end-displacement control, has reactivated in the relevant research fields.

It is known that there are two kinds of basic models for the Euler buckling analysis of beam-like structures, i.e., the force-controlled model and the displacement-controlled model. The former regards the dead forces at the beam ends as the known independent variables and the end displacements as the unknown dependent variables in the corresponding governing equations, while the later considers the end displacements as the given independent variables and the end reaction forces as the unrevealed dependent variables in its governing equations. When we solve a stable equilibrium problem of structures, the force-controlled model is generally considered to be equivalent to the displacement-controlled model, both resulting in the identical results. Hence, the force-controlled model has been prevailingly adopted to investigate the buckling process of beams up to date with almost no attention on its basic difference from the displacement-controlled model (for instance, Holst et al., 2011; Nayfeh and Emam, 2008; Yuan and Wang, 2011; Zhao et al., 2008; Emam and Lacarbonara, 2021; Tissot-Daguette et al., 2022), even when the ends of the structures are actually controlled by the displacement (Su et al., 2012; Cazzolli and Dal Corso, 2019; Cheng and Zhang, 2021; Zhao et al., 2021). Nevertheless, there is a fundamental difference between the two kinds of buckling analysis models if the possible instability during the post-bucking regime is carefully considered (Badiansky, 1974; van der Heijden, 2009; Coulais et al., 2015; Chen and Jin, 2020). The dead end-force may lead a dynamic response to a compressed beam during its post-buckling stage if the prescribed force is greater than the critical buckling force (Badiansky, 1974; Bažant and Cedolin, 2010). Such a situation shall be essentially inconsistent with the static representation of traditional force-controlled model of Euler buckling, and one should generally resort to dynamic model with numerical methods to calculate the complex behavior, as shown by Nayfeh and Emam (2008), Motamarri and Suryanarayan (2012), Alcheikh et al. (2021b) and Hajjaj et al. (2021). Such dynamic results, however, are actually not identical with the real response of the beams under static or quasi-static end-displacement control. Therefore, the force-controlled and displacement-controlled models can no longer be used equivalently in the unstable buckling scenario, and the corresponding model should be chosen according to the actual loading condition.

For the beams under static or quasi-static end-displacement control, it seems that one should solve the governing equations of the displacement-controlled model directly rather than that of the force-controlled model indirectly in view of the difference between the two models. In fact, the energy approach with the asymptotic expansion procedure is traditionally based on the force-controlled model to indirectly calculate the displacement-controlled post-buckling behavior since Koiter's seminal theoretical work (Koiter, 1945; Badiansky, 1974; van der Heijden, 2009; Su et al., 2012; Audoly and Lestringant, 2021). The core of this approach is the determination of an accurate expression of the total potential energy of the post-buckling beam, where the potential energy component of the force has figured out via the product of the formally-known reaction force and the relevant unknown displacement, and the displacement is essentially determined by constructing an approximation of the equilibrium condition with a load factor in the neighborhood of the critical buckling state. If the geometry or deformation of the post-buckling body is a little more complex, the calculation of the energy is practicable but pretty difficult, as shown by van der Heijden (2009), Lubbers et al. (2017), Chen and Jin (2020) and Zhou et al. (2023). On the other hand, however, it seems that no attention has paid to the displacement-controlled model to establish a corresponding calculation procedure in the buckling and post-buckling analysis, even for the simplest Euler buckling problem with the given end-displacement and unknown reaction force.

Here we provide a simple calculation procedure for the analysis of buckling and post-buckling response directly based on the displacement-controlled model without tedious asymptotic expansion of the potential energy. First, a proper buckling displacement field with undetermined coefficients is assumed for the buckling structure according its buckling modes; Second, the total potential energy function is integrated completely based on the displacement-controlled model with the displacement; Third, the coefficients are solved by minimizing the total potential energy function; Fourth, the total potential energy function as well as the buckling displacement are determined by putting back the coefficients; Finally, the internal forces and other related response quantities are deduced via the determined potential energy expression. The five-step procedure is demonstrated via the analytical revisitation of the classical Euler buckling problem, where the explicit expression of the total potential energy in terms of end displacement rather than axial force is first deduced for the buckling and post-buckling beams with various boundary conditions, the solutions to the deformation and internal forces are then obtained in closed-form. These explicit expressions could quantify the Euler buckling and post-buckling process in a more convenient way than the traditional counterparts. Seeing the efficiency in the demonstration, we hope that the simple procedure will be useful in other post-buckling analysis problems in the near further.

## 2. Models and Derivation

### 2. 1. Cosine-shaped buckling

As shown in Fig. 1, we consider a slender elastic beam of initial length $l_0$, cross section area $A$ and second area moment $I$, and Young's modulus $Y$. The beam has two fixed ends, and one of the two displaces statically towards the other one with a value Δ, called the axial end-compression. During the compression, the beam may hold straight or be curved, no matter how it deforms its arc length becomes as $l$ and its projection length on the x-axis is $u$.

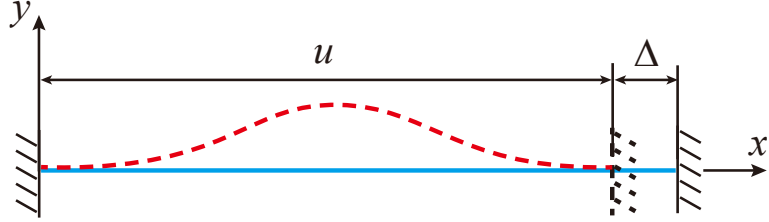

**Fig. 1.** A fixed-fixed slender elastic beam with one end controlled by the axial compression

Taking the beam as an Euler-Bernoulli beam, we initially assume that its deflection curve is of a cosine function satisfied the fixed-fixed boundary condition according to Euler buckling theory (Jerath, 2021),

$$v(x) = \delta(1 - \cos\frac{2\pi x}{u}), (1)$$

where $\delta$ is an undetermined deflection parameter and $(1 - \cos\frac{2\pi x}{u})$ is the buckling mode. Without loss of generality, here the buckling mode can be obtained by the classical theoretical analysis for the modes (Bažant and Cedolin, 2010), and other functions can be used instead of Eq. (1) provided that the functions have a similar shape and satisfy the boundary condition, such as $\delta(2/u)^4 x^2(u-x)^2$ or $\delta(2/u)^6 x^3(u-x)^3$.

Using the deflection function, the length of the deformed beam is obtained as

$$l = \int_0^u \sqrt{1 + v(x)'^2}\, dx = \frac{2u}{\pi} E\left(-\frac{4\pi^2 \delta^2}{u^2}\right), (2)$$

where $E$ denotes the complete elliptic integral of the second kind. Note that it is more convenient to use the following approximation formula to calculate the length (Bažant and Cedolin, 2010),

$$l \approx \int_0^u (1 + v(x)'^2/2)\, dx. (3)$$

We must point out that the length calculation is based on the accompanying configuration of the deformed beam, where the small change of the integral interval is precisely considered with the relation $u = l_0 - \Delta$.

Next, we apply the energy method to determine the unknown parameter $\delta$. In general, the total potential energy $\Pi$ of the beam is the sum of its strain energy $U$ and the work $W$ done by its external force (Bažant and Cedolin, 2010; van der Heijden, 2009), denoted as $\Pi = U - W$. The potential energy of the external force is the work done by the dead "external force" from the "zero point" of potential energy (initial state) to the current deformation position (Reddy, 2002), in other words, the dead force potentially continues to do work on the kinematically admissible displacement field (Anand and Govindjee, 2020). In the traditional force-controlled model, the corresponding displacement is kinematically possible, while in the present displacement-controlled model, the reaction forces at the ends have no such potential to work because the end constraints remain and prevent the kinematically admissible displacement. Therefore, the potential energy of the "external force" is zero for the end-displacement controlled beam shown in Fig. 1. Then, the beam's total potential energy only includes the strain energy $U$, namely $\Pi = U$.

The beam may shorten and bend simultaneously during the compression. Hence, the strain energy of the deformed beam is the sum of compression and bending energy components, which can be calculated by the following equation,

$$U = \frac{YA(l_0-l)^2}{2l_0} + \frac{YI}{2}\int_0^u v(x)''^2 dx, \quad (4)$$

where the first term on the right is the compression strain energy of the beam obtained via the simplification that the axial force is uniform along the beam (Holst et al., 2011; Wang and van der Heijden 2020), and the second term is the bending strain energy of the beam (Bažant and Cedolin, 2010).

Substituting Eq. (1) and Eq. (2) into Eq. (4), we obtain the strain energy and total potential energy of the beam as

$$\Pi = U = \frac{YA}{2l_0}\left(l_0 - \frac{2u}{\pi}E\left(\frac{-4\pi^2\delta^2}{u^2}\right)\right)^2 + \frac{4\pi^4 YI\delta^2}{u^3}, \quad (5)$$

which can be regarded as an energy surface of two generalized coordinates with the relationship $u = l_0 - \Delta$, i.e., the axial end-compression $\Delta$ and deflection parameter $\delta$.

According to the minimum potential energy principle, there must be $\frac{d\Pi(\delta)}{d\delta} = 0$ and $\frac{d^2\Pi(\delta)}{d\delta^2} > 0$ in the stable equilibrium state for the deformed beam. If we depict the state on the potential energy surface, it is the valley bottom path in the energy landscape. However, it's hard to analytically solve the equations to find the exact explicit expression of the path. Approximating the second kind elliptic integral in Eq. (5) by its quadratic Taylor expansion around $\frac{\delta}{u} = 0$, namely $\frac{\pi}{2}(1 + \frac{\pi^2 \delta^2}{u^2})$, the beam's potential energy then becomes

$$\Pi \approx \frac{YA}{2l_0}(l_0 - u - \frac{\pi^2\delta^2}{u})^2 + \frac{4\pi^4 YI\delta^2}{u^3} \approx \frac{YA}{2l_0}(\Delta - \frac{\pi^2\delta^2}{l_0-\Delta})^2 + \frac{4\pi^4 YI\delta^2}{(l_0-\Delta)^3}. \quad (6)$$

Interestingly, the identical approximation of the potential energy is obtained if the length approximation Eq. (3) is employed in Eq. (4).

Solving $\frac{d\Pi(\delta)}{d\delta} = 0$ via Eq. (6), now the deflection parameter is determined, whose solutions shown below are corresponding to the beam's straight and post-buckling states,

$$\delta = 0 \text{ and } \delta = \pm\sqrt{\frac{u(l_0-u)}{\pi^2} - \frac{4Il_0}{Au}} = \pm\sqrt{\frac{u(l_0-u)}{\pi^2} - \frac{4l_0 R_g^2}{u}}, \quad (7)$$

where $R_g = \sqrt{\frac{I}{A}}$, the gyradius of cross section.

Substituting the parameter $\delta$ determined by Eq. (7) into Eq. (5) or (6), we completely express the potential energy function for the beam under the end-compression $\Delta$. Once given the energy function, the solutions to the deformation and internal forces of the beam are consequently deduced from it. To express these results explicitly, we need the critical end-displacement $\Delta_c$ of the buckling onset yet.

Let $\sqrt{\frac{u(l_0-u)}{\pi^2} - \frac{4l_0 R_g^2}{u}} = 0$, we then find a bifurcation point $u_b$. However, its expression is verbose. For convenience, we approximate the bifurcation value $u_b$ as

$$u_b \approx \sqrt{l_0^2 - 8\pi^2 R_g^2} \approx l_0 - \frac{4\pi^2 R_g^2}{l_0}. \quad (8)$$

Substituting the straight state branch $\delta = 0$ into $\frac{d^2\Pi(\delta)}{d\delta^2} = 0$, a valley-ridge inflection point is verified at the bifurcation position $u_b$ along $\delta = 0$ on the energy surface. We calculate $\frac{d^2\Pi(\delta)}{d\delta^2}$ along all the three branches and confirm that the results are greater than zero when $u > u_b$ along the straight branch and $u < u_b$ along the buckled branches. Therefore, the minimum energy path of the beam under the end-compression is first along the straight branch and then along one of the buckled branches. Thus, the point ($\delta = 0$, $u = u_b$) on the potential energy surface is a critical point, which is corresponding to the critical axial end-compression for the Euler buckling behavior,

$$\Delta_c = l_0 - u_b \approx l_0 - \sqrt{l_0^2 - 8\pi^2 R_g^2} \approx \frac{4\pi^2 R_g^2}{l_0}. \quad (9)$$

We then know that the beam is straight and its total potential energy is $\frac{YA\Delta^2}{2l_0}$ before buckling, while it is curved and its total potential energy is $\frac{4\pi^2 YI\Delta}{(l_0-\Delta)^2} - \frac{8\pi^4 l_0 YI^2}{A(l_0-\Delta)^4}$ during post-buckling. The two energy expressions show that with the increase of the end-compression $\Delta$ the energy first increases quadratically and then increase almost linearly, where there is the critical potential energy $\Pi_c \approx \frac{8\pi^4 YI R_g^2}{l_0^3}$.

Now substituting Eq. (7) into Eq. (1) and using $u = l_0 - \Delta$, we completely obtain the beam's deflection curve,

$$v(x) = \begin{cases} 0, & \Delta \leq \Delta_c \\ \pm\sqrt{\frac{\Delta(l_0-\Delta)}{\pi^2} - \frac{4l_0 R_g^2}{l_0-\Delta}}\left(1 - \cos\frac{2\pi x}{l_0-\Delta}\right), & \Delta > \Delta_c \end{cases}. \quad (10)$$

Then the corresponding length is

$$l = \begin{cases} l_0 - \Delta, & \Delta \leq \Delta_c \\ l_0 - \frac{4\pi^2 l_0 R_g^2}{(l_0-\Delta)^2}, & \Delta > \Delta_c \end{cases}. \quad (11)$$

For engineering application, the axial force and bending moment are presented according to the two deformation results,

$$F_N = \begin{cases} \frac{YA\Delta}{l_0}, & \Delta \leq \Delta_c \\ \frac{4\pi^2 YI}{(l_0-\Delta)^2}, & \Delta > \Delta_c \end{cases}, \quad (12a)$$

$$M = \begin{cases} 0, & \Delta \leq \Delta_c \\ \frac{4\pi^2 YI}{(l_0-\Delta)^2}\sqrt{\frac{\Delta(l_0-\Delta)}{\pi^2} - \frac{4l_0 R_g^2}{l_0-\Delta}}\cos\frac{2\pi x}{l_0-\Delta}, & \Delta > \Delta_c \end{cases}. \quad (12b)$$

Eq. (12a) indicates that the axial force of the beam during post-buckling is nearly a constant and equals to the critical reaction force if the deformation is small. Hence, the post-buckling of the beam is approximately a neutral instability in a finite neighborhood.

Finally, substituting Eq. (9) into Eq. (12a) gives two solutions to the critical axial force of the beam,

$$F_{Nc} \approx \begin{cases} \frac{4\pi^2 YI}{l_0^2} \\ \frac{4\pi^2 YI}{l_0^2 - 8\pi^2 R_g^2} \end{cases}. \quad (13)$$

The two expressions in Eq. (13) are similar, where the first one is identical with the classical Euler load formula, while the second one reveals that the critical force also depends on the gyradius of cross section. However, the dependence is very small because $8\pi^2 R_g^2$ is usually much smaller than $l_0^2$.

At this point, a complete displacement-controlled calculation procedure for buckling and post-buckling analysis has been demonstrated by the theoretical analysis of the fixed-fixed beam shown in Fig. 1. Now we further consider the fixed-guided and fixed-free beams under the axial end-displacement control via the procedure. We first assume that these beams have the cosine shape buckling deflection as well (Jerath, 2021),

$$v(x) = \delta(1 - \cos\frac{\mu\pi x}{u}), \quad (14)$$

where the parameter $\mu$ is the shape coefficient, and the fixed-fixed, fixed-guided and fixed-free beams' shape coefficients are 2, 1 and 0.5, respectively. Here note that the deflection amplitude orderly equals $2\delta$, $2\delta$ and $\delta$ for the three cases.

Similar to the above derivation, we obtain the related analytical results. The potential energy surface of the axial end-displacement controlled beams with the shape coefficient $\mu$ is

$$\Pi = \frac{YA}{2l_0}(l_0 - \frac{2u}{\pi}E(\frac{-\mu^2\pi^2\delta^2}{u^2}))^2 + \frac{\mu^4\pi^4 YI\delta^2}{4u^3} \approx \frac{YA}{2l_0}(\Delta - \frac{\mu^2\pi^2\delta^2}{4(l_0-\Delta)})^2 + \frac{\mu^4\pi^4 YI\delta^2}{4(l_0-\Delta)^3}. \quad (15)$$

The deflection parameter is deduced as

$$\delta = 0 \text{ and } \delta = \pm\sqrt{\frac{4\Delta(l_0-\Delta)}{\mu^2\pi^2} - \frac{4Il_0}{A(l_0-\Delta)}} \approx \pm\sqrt{\frac{4\Delta l_0}{\mu^2\pi^2} - \frac{4l_0 R_g^2}{l_0-\Delta}}, \quad (16)$$

where the approximation $l_0 \approx l_0 - \Delta$ can be used for the small deformation.

The corresponding deformation curve is

$$v(x) = \begin{cases} 0, & \Delta \leq \Delta_c \\ \pm\sqrt{\frac{4\Delta(l_0-\Delta)}{\mu^2\pi^2} - \frac{4l_0 R_g^2}{l_0-\Delta}}(1 - \cos\frac{\mu\pi x}{l_0-\Delta}), & \Delta > \Delta_c \end{cases}, \quad (17)$$

and the curve length is

$$l = \begin{cases} l_0 - \Delta, & \Delta \leq \Delta_c \\ l_0 - \frac{\mu^2\pi^2 l_0 R_g^2}{(l_0-\Delta)^2}, & \Delta > \Delta_c \end{cases}, \quad (18)$$

where the critical position and the critical axial end-compression are

$$u_b \approx \sqrt{l_0^2 - 2\mu^2\pi^2 R_g^2}, \quad (19a)$$

$$\Delta_c \approx l_0 - \sqrt{l_0^2 - 2\mu^2\pi^2 R_g^2} \approx \frac{\mu^2\pi^2 R_g^2}{l_0}. \quad (19b)$$

The minimum potential energy for the deformed beam is obtained as

$$\Pi_m = \begin{cases} \frac{YA\Delta^2}{2l_0}, & \Delta \leq \Delta_c \\ \frac{\mu^2\pi^2 YI\Delta}{(l_0-\Delta)^2} - \frac{\mu^4\pi^4 l_0 YI^2}{2A(l_0-\Delta)^4}, & \Delta > \Delta_c \end{cases}. \quad (20)$$

The axial force and bending moment are

$$F_N = \begin{cases} \frac{YA\Delta}{l_0}, & \Delta \leq \Delta_c \\ \frac{\mu^2\pi^2 YI}{(l_0-\Delta)^2}, & \Delta > \Delta_c \end{cases}, \quad (21a)$$

$$M = \begin{cases} 0, & \Delta \leq \Delta_c \\ \frac{\mu^2\pi^2 YI}{(l_0-\Delta)^2}\sqrt{\frac{4\Delta(l_0-\Delta)}{\mu^2\pi^2} - \frac{4l_0 R_g^2}{l_0-\Delta}}\cos\frac{\mu\pi x}{l_0-\Delta}, & \Delta > \Delta_c \end{cases}. \quad (21b)$$

The critical axial force is

$$F_{Nc} \approx \frac{\mu^2\pi^2 YI}{l_0^2} \approx \frac{\mu^2\pi^2 YI}{l_0^2 - 2\mu^2\pi^2 R_g^2}. \quad (22)$$

The corresponding critical axial stress, axial strain and potential energy can be simply given by $\sigma_c = F_{Nc}/A$, $\varepsilon_c = \Delta_c/l_0$ and $\Pi_c = \frac{YA}{2l_0}\Delta_c^2$, respectively.

We further normalize the total potential energy Eq. (15) by the critical energy. A dimensionless energy function with only two dimensionless variables $\Delta/\Delta_c$ and $\frac{\delta}{R_g}$ is given,

$$\frac{\Pi}{\Pi_c} \approx \left(\frac{\Delta}{\Delta_c} - \frac{1}{4}(\frac{\delta}{R_g})^2\right)^2 + \frac{1}{2}(\frac{\delta}{R_g})^2, (23)$$

and the corresponding normalized minimum potential energy depends on only one variable $\Delta/\Delta_c$, i.e.,

$$\frac{\Pi_m}{\Pi_c} = \begin{cases} \frac{\Delta^2}{\Delta_c^2}, & \frac{\Delta}{\Delta_c} \leq 1 \\ \frac{2\Delta}{\Delta_c} - 1, & \frac{\Delta}{\Delta_c} > 1 \end{cases}. (24)$$

Interestingly, the normalized potential energy surface and its extrema trajectories are universal for all the three types of beams. Meanwhile, Eq. (24) confirms the energy's quadratic dependence on the axial end-compression before buckling and its linear dependence after the buckling.

## 2. 2. Sine-shaped and mixed buckling

Here we reconsider the Euler buckling behavior of the pinned-pinned, pinned-guided and fixed-pinned beams under the axial end-displacement control via the displacement-controlled analysis procedure. The pinned-pinned and pinned-guided beams are assumed to buckle in the following sine shape (Jerath, 2021),

$$v(x) = \delta \sin\frac{\mu\pi x}{u}, (25)$$

where the shape coefficient $\mu$ equals 1 and 0.5 for the for the pinned-pinned and pinned-guided beam respectively. Here please note that the deflection amplitude equals to the deflection parameter $\delta$ for the two cases.

Similar to the above deduction of the cosine-shaped buckling situation, the potential energy of the sine-shaped beams with the axial end-compression can be calculated via Eq. (4) with Eq. (2) and Eq. (25). However, the energy result consequently yields tedious solutions to the deflection parameter $\delta$, critical position and so on. The curve length approximation Eq. (3) is then used instead of Eq. (2). We have obtained the corresponding potential energy expressions of the beams and find that the pinned-pinned beam has the same potential energy as the fixed-guided beam, and the pinned-guided beam has the same potential energy as the fixed-free beam. Consequently, the deflection parameter, curve length, critical position, axial force, and critical loads of the pinned-pinned beam are respectively identical to that of the fixed-guided beam, and these results of the pinned-guided beam are respectively identical to that of the fixed-free beam. The corresponding deformation curve and bending moment can be readily obtained by replacing the cosine shape in Eq. (17) and Eq. (21b) by the sine shape respectively, i.e.,

$$v(x) = \begin{cases} 0, & \Delta \leq \Delta_c \\ \pm\sqrt{\frac{4\Delta(l_0-\Delta)}{\mu^2\pi^2} - \frac{4l_0 R_g^2}{l_0-\Delta}} \sin\frac{\mu\pi x}{l_0-\Delta}, & \Delta > \Delta_c \end{cases}, (26a)$$

$$M = \begin{cases} 0, & \Delta \leq \Delta_c \\ \frac{\mu^2\pi^2 YI}{(l_0-\Delta)^2}\sqrt{\frac{4\Delta(l_0-\Delta)}{\mu^2\pi^2} - \frac{4l_0 R_g^2}{l_0-\Delta}} \sin\frac{\mu\pi x}{l_0-\Delta}, & \Delta > \Delta_c \end{cases}. (26b)$$

Lastly, we deal with the buckling problem of the fixed-pinned beam under the axial end-displacement control. Its buckling deflection curve is assumed to have the following mixed sine shape (Jerath, 2021), whose deflection amplitude equals $\mu\delta$,

$$v(x) = \delta(\frac{x}{u} - \sin\frac{\mu\pi x}{u} / \sin\mu\pi). (27)$$

Through the similar displacement-controlled derivation, we find that the potential energy, deflection parameter, curve length, critical position, axial force and critical loads of the fixed-pinned beam can be calculated by the Eq. (15), Eq. (16), Eq. (18), Eq. (19), Eq. (21a) and Eq. (22) with $\mu \approx 1.43$, respectively. Replacing the sine shape function in Eq. (17) and Eq. (21b) by the mixed shape function gives the deformation curve and bending moment for the fixed-pinned beam,

$$v(x) = \begin{cases} 0, & \Delta \leq \Delta_c \\ \pm\sqrt{\frac{4\Delta(l_0-\Delta)}{\mu^2\pi^2} - \frac{4l_0 R_g^2}{l_0-\Delta}} \left(\frac{x}{l_0-\Delta} - \frac{\sin\frac{\mu\pi x}{l_0-\Delta}}{\sin\mu\pi}\right), & \Delta > \Delta_c \end{cases}, (28a)$$

$$M = \begin{cases} 0, & \Delta \leq \Delta_c \\ \frac{\mu^2\pi^2 YI}{(l_0-\Delta)^2 \sin\mu\pi}\sqrt{\frac{4\Delta(l_0-\Delta)}{\mu^2\pi^2} - \frac{4l_0 R_g^2}{l_0-\Delta}} \sin\frac{\mu\pi x}{l_0-\Delta}, & \Delta > \Delta_c \end{cases}. (28b)$$

So far, we have completed the buckling and post-buckling analysis for the beams with the six typical displacement-controlled boundary conditions. We can further imagine a universal boundary constraint that could steadily control the beam end from the completely fixed state to the smooth hinge state. With such constraints the shape coefficient $\mu$ should generally has any values in the range from 0.5 to 2, and we can use the above analytical solutions to predict the corresponding beams' buckling behavior, including their potential energy, deflection parameter, curve length, critical position, axial force and critical loads. Hence, the effect of the boundary conditions is quantified by a proper value of the shape coefficient, which is interestingly the reciprocal of the effective length factor used in the well-known Euler load formula.

## 3. Result illustration

Here we show the analytical results of the displacement-controlled procedure with the comparison with the counterparts of the traditional force-controlled procedure. We first display the universal dimensionless potential energy surface Eq. (23) in Fig. 2 (a) for the compressed elastic beams with various boundary conditions, while whose counterpart of the force-controlled model has no closed-form yet. The energy surface has only one unique shape depending on the two dimensionless quantities, $\Delta/\Delta_c$ and $\delta/R_g$. The extrema trajectories of the surface are also plotted in the graph via Eq. (24). With the increasing axial end-compression, the potential energy first increases along the minima pathway $\delta/R_g = 0$, then inflect-bifurcates aside into the one of the two minima pathways $\delta/R_g = \pm\sqrt{\frac{4\Delta}{\Delta_c} - 4}$ at the critical point $\frac{\Delta}{\Delta_c} = 1$, and the original pathway $\delta/R_g = 0$ becomes the maxima pathway beyond the point. It is noted that the initial straight beams buckle into one side completely randomly at the critical point, where the beams' symmetry breaks spontaneously.

Three potential wells nearby the critical state are shown in Fig. 2 (b), where $\Delta = \Delta_c$, $\Delta = 0.9\Delta_c$ and $\Delta = 1.1\Delta_c$. We see that there is one flat bottom for the case $\Delta = \Delta_c$, one sunken bottom for the case $\Delta = 0.9\Delta_c$ and two sunken bottoms for the case $\Delta = 1.1\Delta_c$. Disregarding the slight difference, their bottoms ranging from about -1 to 1 are almost flat. That suggests that the beams deflect very easily and are very sensitive to small disturbance in the region, in other words, the beams have a nearly zero bending stiffness near the critical state. Fig. 2 (c) further shows the normalized minimum potential energy Eq. (24). Its universal dependence on the dimensionless axial end-compression is quadratic before buckling and linear with a slope of 2 after buckling.

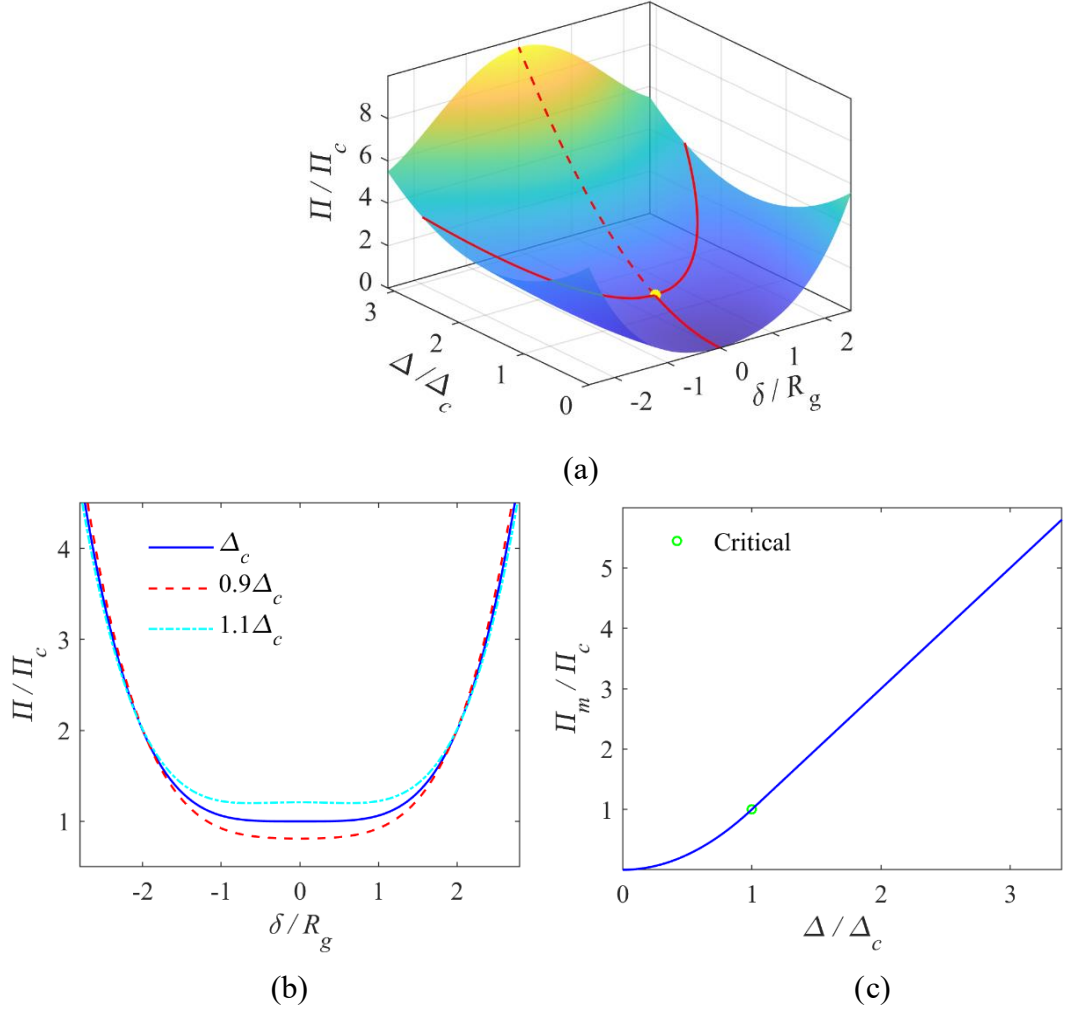

**Fig. 2.** (a) Universal dimensionless potential energy surface as a function of the dimensionless axial end-compression and deflection parameter for the compressed beams with various boundary conditions, where the solid line is the minima path, the dashed line is the maxima path, and the yellow dot denotes the bifurcation-inflection point. (b) Three potential wells nearby the bifurcation-inflection point. (c) The dependence of dimensionless minimum potential energy on the dimensionless axial end-compression.

Besides the abstract universal energy function, we next show the more practical application of our other analytical solutions. The steel beam-like specimens tested by Carpinteri et al. (2014) is considered to illustrate our analytical solutions with four representative shape coefficient values, 0.5, 1.0, 1.43 and 2.0, whose details are listed in Table. 1.

**Table. 1.**

Size, mechanical property and Euler loads of the steel beam-like specimens under the axial end-compression tested by Carpinteri et al. (2014).

| Specimen | Length (mm) | Section (mm$^2$) | Straightness deviation | Elasticity (GPa) | Bucking load (kN) |
|---|---|---|---|---|---|
| Pinned-pinned | 1000 | 15×30 | <0.2/1000 | 210 | 17.6 |
| Fixed-pinned | 1000 | 15×30 | <0.2/1000 | 210 | 30.0 |

Eq. (20) indicates that the potential energy of the deformed beams has a quadratic dependence on the axial end-compression before buckling and has a complex dependence after buckling. However, Fig. 3 (a) demonstrates that if the deformation is finite, not large than the moderate deflection, there is an almost linear dependence on the axial end-compression during post-buckling, where the critical values of $\Delta_c$ and $\Pi_c$ are (0.046mm, 0.101J), (0.185mm, 1.618J), (0.378mm, 6.769J) and (0.740mm, 25.91J) in turn for $\mu$ =0.5, 1.0, 1.43 and 2.0. We guess that the energy transition from the quadratic to the linear is a key character for other structures' instability behavior as well.

We imagine an almighty end constraint at one end of the beams, whose shape coefficient could equal to any value in the range (0.5~2). The influence of the constraint is plotted in Fig. 3 (b) via three representative cases with different axial end-compressions, 0.2mm, 0.4mm and 0.6mm. The shape coefficient's effect on the minimum potential energy is divided into two stages by the beams' critical states: the increasing shape coefficient nonlinearly increases the minimum energy in the left part, whereas the coefficient has no influence on the energy in the right part.

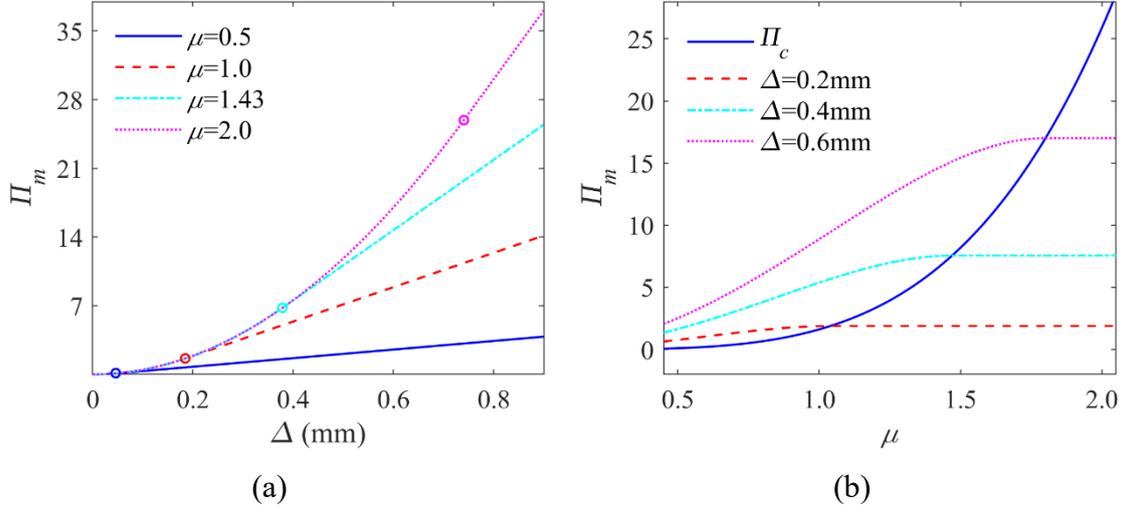

(a)                      (b)

**Fig. 3.** Influence of the axial end-compression (a) and shape coefficient (b) on the minimum potential energy of the beams under axial end-compression, calculated by Eq. (20) with values in Table 1, where the circles denote the critical states.

The deformation results, including the deflection and shortening, are plotted in Fig. 4 for the buckling beams under the axial end-displacement control. By increasing the axial end-compression, sudden turns occur in their deflection and length, where the critical end-compressions are 0.046mm, 0.185mm, 0.378mm and 0.740mm for the beams with $\mu$ =0.5, 1.0, 1.43 and 2.0, respectively. Before the turning points, the deflection parameter keeps zero while the length decreases linearly, which implies that the axial shortening dominates the beams' deformation before buckling. After the turning points, the deflection parameter increases nonlinearly while the length keeps constant, which suggests the bending increasingly dominates the deformation during post-buckling. Besides, the beams with bigger shape coefficients have a bigger deflection parameter but a shorter length for a same axial end-compression after buckling. Here we should note that the deflection parameter $\delta$ is not the deflection amplitude, and their difference is a constant determined by the post-buckling shape function.

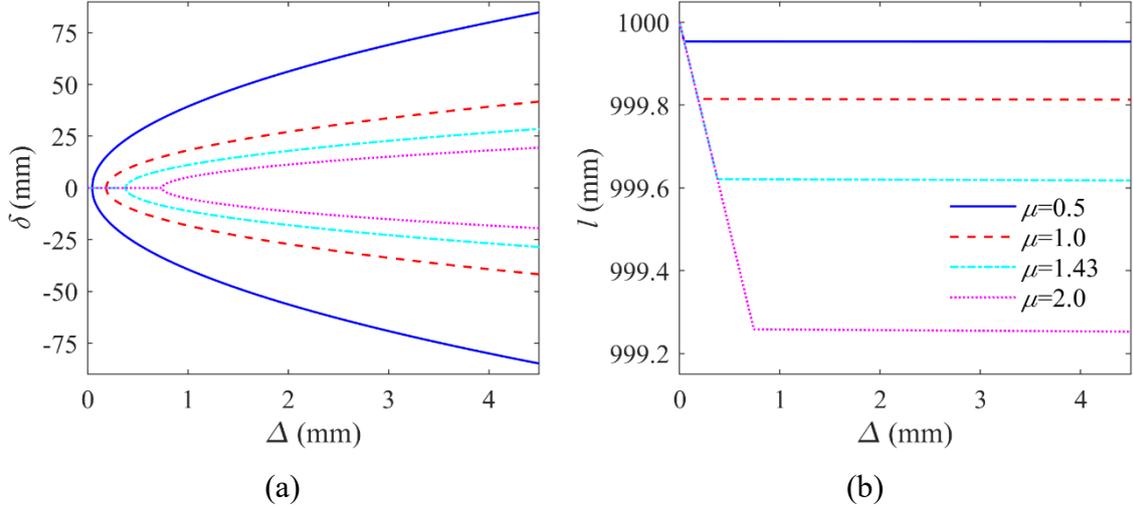

(a) (b)

**Fig. 4.** Deflection parameter (a) and length (b) of the beams under axial end-compression with different shape coefficients, calculated by Eq. (16) and Eq. (18) with values in Table 1.

The beams' internal forces (axial force and maximum bending moment) display similar sudden turns, as shown in Fig. 5, where $\Delta_c$ equals 0.046mm, 0.185mm, 0.378mm and 0.740mm for $\mu =0.5$, 1.0, 1.43 and 2.0 in turn. We also plot the counterparts of the load-shortening relation by the traditional force-controlled model in the Fig. 5a, exactly showing a small discrepancy between the two kinds of results of the displacement-controlled and force-controlled models. It is clear that the axial forces initially increase linearly with the axial end-compression, while the maximum bending moments equal to zero before buckling. During post-buckling, the axial forces have a slight growth with the increasing end-compression, almost remaining at the critical axial forces, while the maximum bending moments fast increase with the end-displacement especially after just buckling, and the bigger the shape coefficient is, the steeper the increase is.

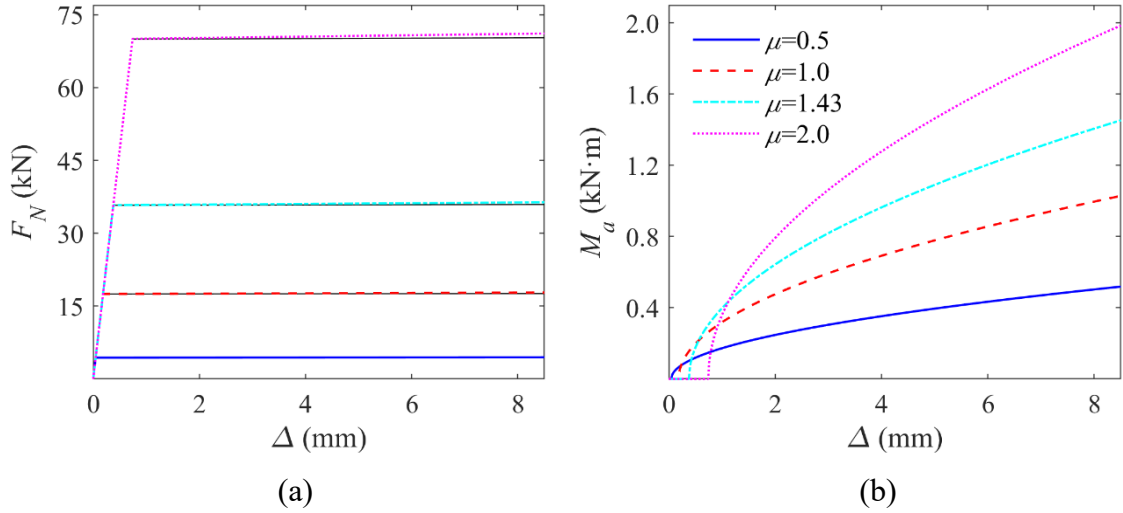

**Fig. 5.** (a) Axial forces of the beams under axial end-compression with different shape coefficients, calculated by Eq. (21a) with values in Table 1 (colored lines), and by its counterpart of the force-controlled model (black lines) whose post-buckling slope is 1/2 (van der Heijden, 2009; Coulais et al., 2015). (b) The corresponding maximum bending moments of the beams, calculated by Eq. (21b).

Fig. 6 further demonstrates the axial forces of the beams as a function of the dimensionless deflection parameter. The curves for the beams with various shape coefficients show clearly that the bifurcations of the beams are almost symmetric neutral bifurcation. The neutral bifurcation implies that a certain end-force of the beams corresponds an uncertain deflection amplitude of many values during the post-buckling. It is neutral bifurcation that causes the long-standing challenge in the theoretical analysis of post-buckling behavior via the traditional force-controlled model.

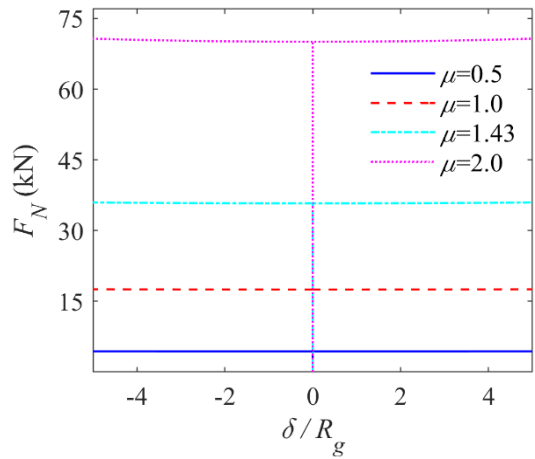

**Fig. 6.** Axial forces as a function of the deflection parameter for the beams under the axial end-compression with different shape coefficients, calculated by Eq. (21) and Eq. (16) with values in Table 1.

Lastly, we examine in Fig. 7 the critical loads of the buckling beams under the axial end-compression to see the influence of the beam section gyradius and shape coefficient on the critical end-compressions and the corresponding stresses. The critical axial end-compressions are determined by Eq. (19b) for the beams with four shape coefficients 0.5, 1.0, 1.43 and 2.0, which depend only on the beams' geometric factors. The stress results are calculated by the second expression in Eq. (22). Meanwhile the corresponding results of the Euler buckling load formula are also denoted by the solid dots in Fig. 7a, with an excellent agreement. We see a quadratic dependence of the buckling end-compression or stress on the gyradius. In Fig. 7b, the critical loads of the beams with three gyradii values indicate that the increasing shape coefficient rises the critical loads in a gradual way, and the bigger the gyradius is, the steeper the growth is.

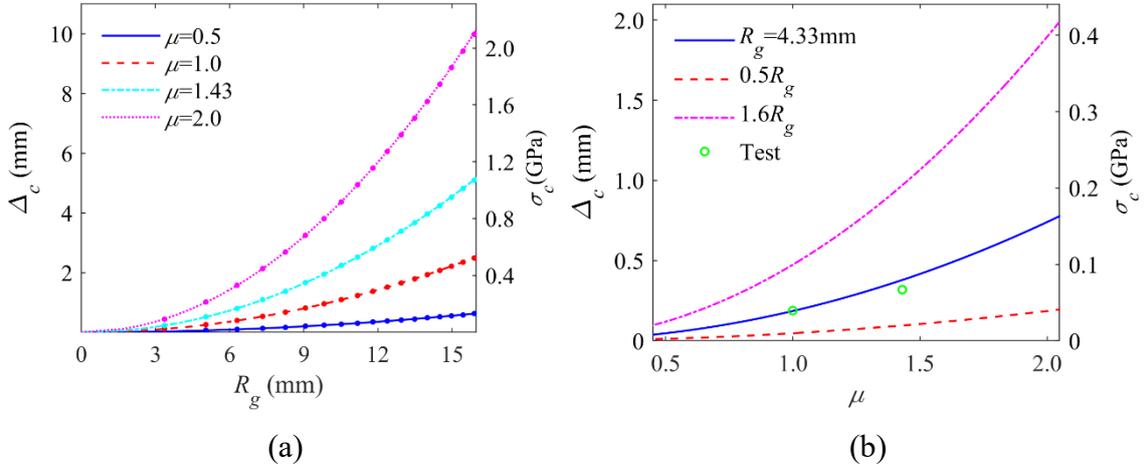

(a)        (b)

**Fig. 7.** Influence of the beam section gyradius (a) and shape coefficient (b) on the critical axial end-compression and stress, calculated by Eq. (19) and Eq. (22) with $\sigma_c = F_{Nc}/A$ and the values in Table 1, where the solid dots denote the corresponding results by the Euler buckling force formula, the test results are from the work by Carpinteri et al. (2014).

## 4. Conclusions

In this work, we show a post-buckling analysis procedure to directly solve the buckling problem under displacement control and to avoid the energy asymptotic expansion of the traditional force-controlled approach, via the example of the classical Euler buckling problem with various boundary conditions. In the simple procedure, the beams' total potential energy function is precisely determined on the deformed configuration in terms of end-displacement rather than the end-force, where the axial compression and bending energy components are both took into account for the total potential energy of the beams, but no the potential energy component of the dead force. Analytical solutions to the deformation and internal forces during the whole finite buckling process are consequently obtained for the beams. In comparison with the counterpart solutions of the traditional force-controlled model, our closed-form results directly express the Euler buckling and post-buckling responses in a convenient way. Besides some rediscovery of the traditional results, the following conclusions with new details are drawn for the beams under the axial end-compression:

(1) The beams with various boundary conditions have a unique but universal dimensionless potential energy surface, whose minima pathways have a quadratic dependence on the axial end-compression before buckling and an almost linear dependence with a slope of 2 after buckling;

(2) The beams experience almost symmetric neutral bifurcation buckling when a small discrepancy in the load-shortening relation exists between the two kinds of results of the displacement-controlled and force-controlled models, whose post-buckling slope is slightly greater than 1/2.

For the engineering buckling and post-buckling analysis of end-displacement controlled beams, our closed-form solutions are ready to predict the beams' deformation profile, axial force, bending moment, critical axial end-compression throughout the whole finite buckling process. Besides the beam-like structures, the present displacement-controlled analysis procedure can be applied to reexamine the buckling and post-buckling behavior of other structures under displacement control, such as the plates and shells. We believe that new analytical solutions and deeper theoretical understanding will be obtained on the charming buckling problems through the procedure.

**Acknowledgment**

The author acknowledges the financial support from the Natural Science Foundation of Guangdong Province (Grant No. 2019A1515011909).